\documentclass[twocolumn]{aastex63}

\usepackage{graphicx}

\submitjournal{ApJ}

\shorttitle{Multi-instrument tracking of a Type III burst}
\shortauthors{Badman et al.}
\graphicspath{{./}{figures/}}
\begin{document}

\title{Tracking a beam of electrons from the low solar corona into interplanetary space \\ with the Low Frequency Array, Parker Solar Probe and 1 au spacecraft.}

\correspondingauthor{Samuel T. Badman}
\email{samuel\_badman@berkeley.edu}

\author[0000-0002-6145-436X]{Samuel T. Badman}
\affil{Physics Department, University of California, Berkeley, CA 94720-7300, USA}
\affil{Space Sciences Laboratory, University of California, Berkeley, CA 94720-7450, USA}

\author[0000-0002-6106-5292]{Eoin Carley}
\affiliation{Astronomy \& Astrophysics Section, Dublin Institute for Advanced Studies, D02 XF86, Ireland.}

\author[0000-0003-4711-522X]{Luis Alberto Ca\~nizares}
\affiliation{Astronomy \& Astrophysics Section, Dublin Institute for Advanced Studies, D02 XF86, Ireland.}
\affiliation{School of Physics, Trinity College Dublin, Dublin 2, Ireland.}

\author[0000-0003-3903-4649]{Nina Dresing}
\affiliation{Department of Physics and Astronomy, University of Turku, Finland}

\author[0000-0002-6849-5527]{Lan K. Jian}
\affiliation{Heliophysics Science Division, NASA Goddard Space Flight Center, Greenbelt, MD 20771, USA}

\author[0000-0002-3176-8704]{David Lario}
\affiliation{Heliophysics Science Division, NASA Goddard Space Flight Center, Greenbelt, MD 20771, USA}

\author[0000-0001-9745-0400]{Peter T Gallagher}
\affiliation{Astronomy \& Astrophysics Section, Dublin Institute for Advanced Studies, D02 XF86, Ireland.}

\author[0000-0002-2587-1342]{Juan C. Mart\'inez Oliveros}
\affil{Space Sciences Laboratory, University of California, Berkeley, CA 94720-7450, USA}

\author[0000-0002-1573-7457]{Marc Pulupa}
\affiliation{Space Sciences Laboratory, University of California, Berkeley, CA 94720-7450, USA}

\author[0000-0002-1989-3596]{Stuart D. Bale}
\affiliation{Physics Department, University of California, Berkeley, CA 94720-7300, USA}
\affiliation{Space Sciences Laboratory, University of California, Berkeley, CA 94720-7450, USA}

\begin{abstract}
Type III radio bursts are the result of plasma emission from mildly relativistic electron beams propagating from the low solar corona into the heliosphere where they can eventually be detected \textit{in situ} if they align with the location of a heliospheric spacecraft. Here we observe a type III radio burst from 0.1-16~MHz using the Parker Solar Probe (PSP) FIELDS Radio Frequency Spectrometer (RFS), and from 20 - 80 MHz using the Low Frequency Array (LOFAR). This event was not associated with any detectable flare activity but was part of an ongoing type III and noise storm that occurred during PSP encounter 2. A deprojection of the LOFAR radio sources into 3D space shows that the type III radio burst sources were located on open magnetic field from 1.6-3~R$_\odot$  and originated from a near-equatorial active region around longitude E48$^o$. Combining PSP/RFS observations with WIND/WAVES and STEREO/WAVES, we reconstruct the type III radio source trajectory in the heliosphere interior to PSP’s position, assuming ecliptic confinement. An energetic electron enhancement is subsequently detected \textit{in situ} at the STEREO-A spacecraft at compatible times although the onset and duration suggests the individual burst contributes a subset of the enhancement. This work shows relatively small-scale flux emergence in the corona can cause the injection of electron beams from the low corona into the heliosphere, without needing a strong solar flare. The complementary nature of combined ground and space-based radio observations, especially in the era of PSP, is also clearly highlighted by this study.
\end{abstract}



\section{Introduction} 
\label{sec:intro}

Type III radio bursts are the characteristic signature of beams of energetic electrons injected by transient processes in the solar corona onto interplanetary magnetic field (IMF) lines first classified from observations of solar impulsive emission by \citet{Wild1950}. As the electron beams propagate away from the Sun, they produce Langmuir waves ($L$) which mode convert \citep[via one of several possible processes, e.g. ][]{Cairns1987a,Cairns1987b,Melrose2017} to radio emission at approximately the local electron plasma frequency ($f_{pe}$) or its second harmonic. As the electron plasma frequency is proportional to the square root of the electron plasma density ($n_e$) which decreases with the distance from the Sun, type III bursts are readily identified in radio spectrograms as transient signals dropping in frequency with time on a characteristic timescale of minutes to hours \citep{Wild1950}. This frequency ranges from as high as $\mathcal{O}$(1 GHz) very near the solar surface \citep[although typically type III bursts are observed to start at 10s-100s of MHZ in the low corona, ][]{Reid2014}, and can extend down to $\mathcal{O}$(10 kHz) at $\sim$1~au. Above 10-15~MHz, type IIIs are readily observed and imaged by ground based observatories out to radial distances of a few solar radii. At lower frequencies, the burst passes below the typical cut-off frequency of the Earth's ionosphere and therefore can only be detected by space-based radio antennas. A majority of interplanetary type III bursts (where the electron beam escapes onto open field lines) are below this cutoff and can only be observed from space where radio imaging is not currently possible. The interplanetary range is often considered to cover the transition from hectometric to kilometric wavelengths. Therefore, we consider also radio bursts observed at a frequency of 1~MHz (distances of about 7 R$_\odot$) as interplanetary radio bursts.

Further, this radio emission is widely beamed \citep[e.g. ][]{Lecacheux1989,Bonnin2008} and very luminous, meaning emission from the same event can be detected not just by Earth-based observatories, but also by widely separated spacecraft in the inner heliosphere. Additionally, the electron beam itself can survive out beyond 1 au and provide additional \textit{in situ} information of the burst such as near-relativistic electron intensity enhancements \citep[e.g.][]{Ergun1998} or direct detection of Langmuir waves \citep[e.g. ][]{Gurnett1976}. For further context, \citet{Reid2014} present a relatively recent review on type III burst properties and current theoretical understanding of the phenomena.

Type III radio bursts therefore produce a wide variety of diagnostic information about themselves and the ambient properties of the heliosphere such as the radial evolution of the solar wind plasma density and magnetic topology of the IMF lines \citep{Li2016}. However, much of this information requires obtaining knowledge about the burst's trajectory through the corona and interplanetary space. For example, the electron beam trajectory traces open field lines starting low in the corona all the way out into interplanetary space. They therefore can be a passive tracer of both open coronal magnetic field \citep{Pick2009,Pick2016} and over larger scales in the heliosphere where the magnetic field forms the channels along which space weather events are guided. Additionally, since the emission frequency is directly related to the local plasma frequency and therefore plasma density, an accurate trace of the burst provides a trace through the density structure of the heliosphere allowing testing or development of density models \citep[the latter of which was done by ][]{Leblanc1998}. Further, since emission of type III radio emission occurs near the local plasma frequency, electromagnetic dispersion of the emitted waves is important. This means that propagation of the radio waves is affected by density variations, both through large scale gradients \citep[e.g. ][]{Thejappa2010,Thejappa2011} and by small scale inhomogeneities \citep[e.g. ][]{Li2012,Reid2021}. For example, if radio emission is formed in a localized pocket of reduced plasma density surrounded by overdense regions, the radiation would not propagate through the overdense regions and so it would not be detected by widely separated spacecraft. Smaller scale density fluctuations also lead to random walks of photons before they escape to free stream. Measurements of apparent source size and source height allow these fluctuations to be quantified both in coronal contexts \citep{Kontar2019} and in the heliosphere \citep{Krupar2020,Musset2021}.

In this work we combine multiple observational signatures of a single type III radio burst observed by three well-separated spacecraft as well as ground-based telescopes during the outbound phase of the second solar encounter of NASA's Parker Solar Probe \citep[PSP; ][]{Fox2016} on 2019 April 9. We combine these observations in an effort to track the injection and escape from the corona of the electrons responsible for the type III burst, the interplanetary trajectory of those electrons, and their \textit{in situ} signature at $\sim$1~au. In section \ref{sec:obs} we introduce the various data sources and observations used in the project. The relative positions of these observing spacecraft and instruments are depicted in figure \ref{fig:solar-mach}. In section \ref{sec:methods} we discuss the different methods used to probe the type III's behaviour on its journey out from the sun and the resulting information discovered about its journey from the corona to 1~au. In particular, we focus on ground based radio imaging of the burst in the outer corona (section \ref{subsec:lofar}), time-of-flight analysis between different heliospheric spacecraft to probe the burst's trajectory in the inner heliosphere (section \ref{subsec:TDOA}), and finally study \textit{in situ} signatures of the electron beam arrival at the Solar Terrestrial Relations Observatory A \citep[STEREO A; ][]{Kaiser2005} at 1~au (section \ref{subsec:insitu}). Lastly, in section \ref{sec:discussion} we discuss the results, interpretation and future outlook of our results and future opportunities for collaborative studies between ground and space-based radio observatories and other supporting instrumentation.

\begin{figure*}[t!]
\includegraphics[width=\textwidth]{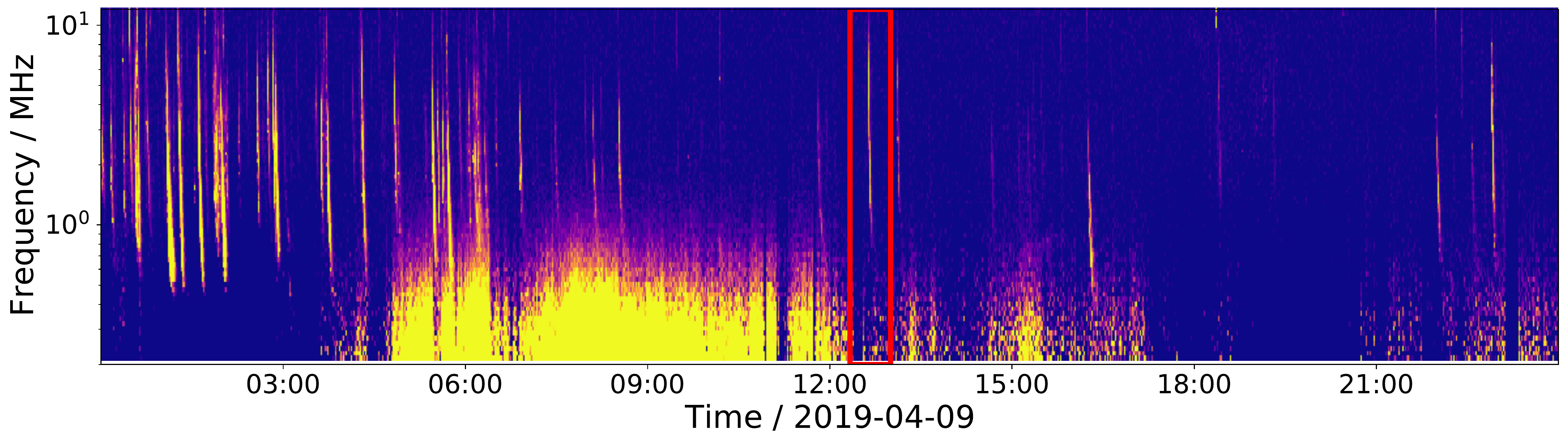}% This is a *.eps file
\caption{PSP/RFS radio spectrogram for 2019 April 9th. Each vertical stripe is a Type III burst. The $\sim$12:40 UT burst studied in this paper is indicated by the red box annotating the figure. The broadband signatures at lower frequencies are due to variations in the in situ electron plasma parameters \citep{Moncuquet2020, Liu2021}}
\label{fig:apr9}
\end{figure*}

\begin{figure}[t!]
\includegraphics[scale=0.16, trim=1cm 1cm 1cm 0cm]{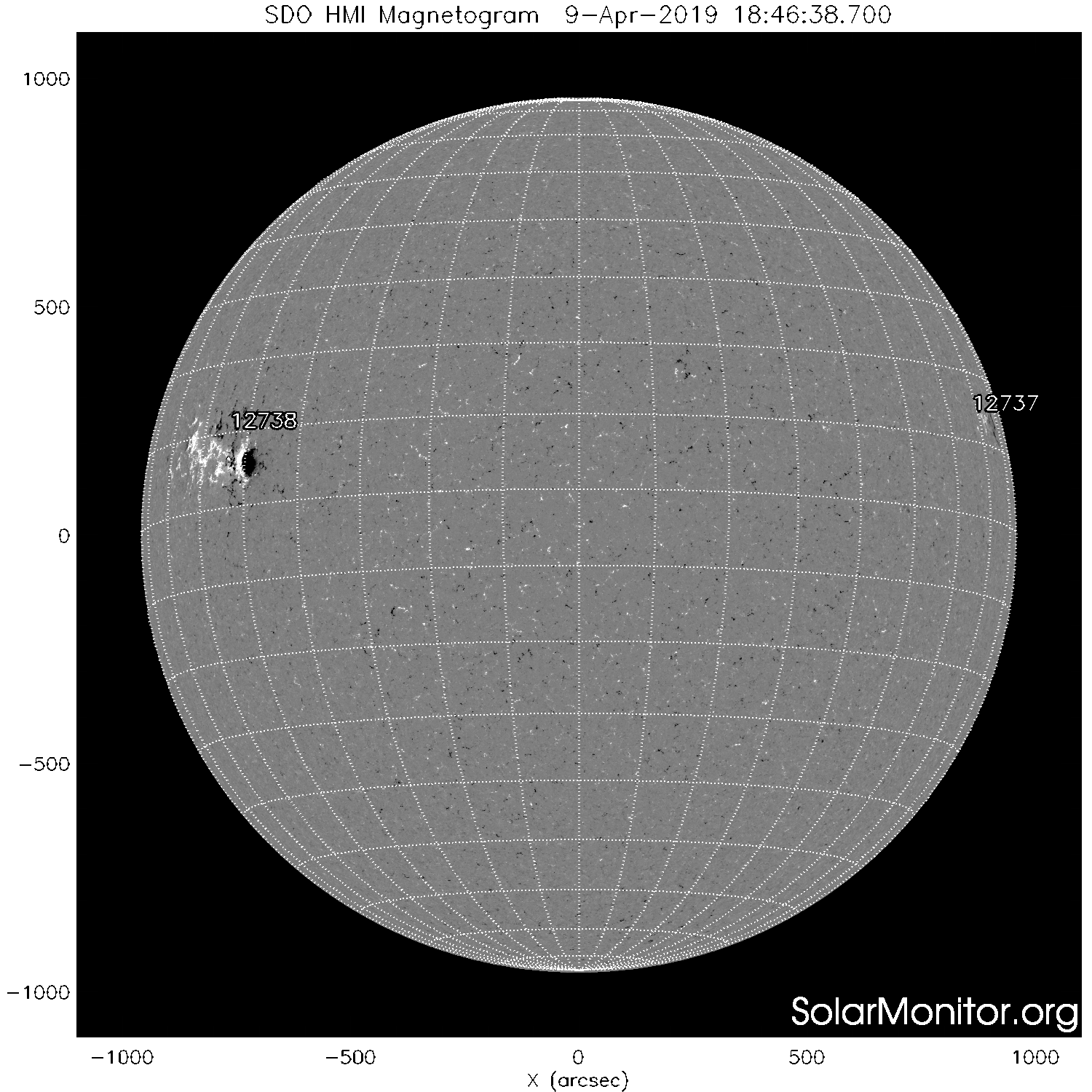}% This is a *.eps file
\caption{HMI magnetogram showing AR 12738 located near the east limb of the sun (48$^o$E). During PSP encounter 2, this AR was associated with a long-duration type III storm and noise storm \citep{Pulupa2020,Krupar2020,Cattell2021}}
\label{fig:hmi}
\end{figure}

\section{Observations and Data Sources} 
\label{sec:obs}

While PSP launched in an extremely quiet solar minimum, its second solar encounter in late March and April 2019 occurred during a wealth of solar radio activity with hundreds of individual type III bursts detected over this interval \citep{Pulupa2020}. During this time, at least two active regions (ARs) were potentially contributing to this activity \citep[AR 12738; ][]{Krupar2020,Cattell2021} and \citep[AR 12737; ][]{Harra2021}. In this work, we focus on a particular, well isolated Type III burst observed on 2019 April 9. The radio context for this day as observed by the FIELDS Radio Frequency Spectrometer \citep[FIELDS/RFS; ][]{Bale2016,Pulupa2016} on board PSP is shown in Figure \ref{fig:apr9} with the individual burst of interest indicated by a red box around 12:40~UT. While the burst is one of many on this day and not the most intense, it is chosen for this study since it is well isolated (a single electron beam) and clearly observed by multiple spacecraft and ground based instrumentation.

\begin{figure}[ht!]
\vspace{2mm}
\includegraphics[scale=0.41, trim=1cm 1cm 10cm 1cm]{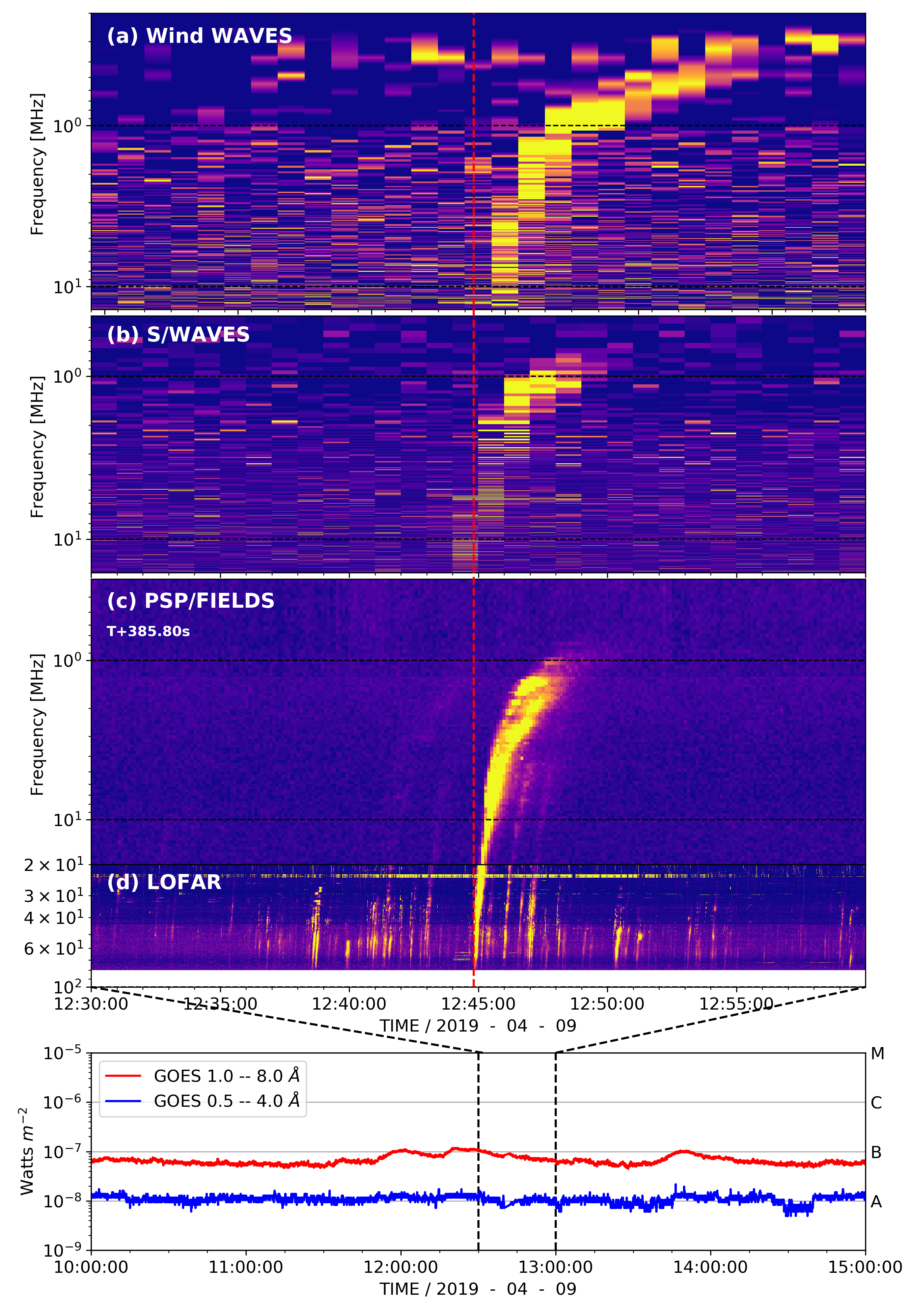}
\caption{A type III radio burst observed with (a) WIND/WAVES, (b) STEREO-A/WAVES (S/WAVES), (c) PSP/FIELDS and (d) LOFAR. The radio burst is observed at LOFAR, S/WAVES and WIND/WAVES at $\sim$12:45~UT as they are all located at a distance of $\sim$1~au from the Sun, while it was observed 6 minutes earlier by PSP due to its closer proximity to the sun at $\sim 50 R_\odot$. Using a PSP-Earth light travel time, PSP's dynamic spectra has been shifted by +385.80s to match the one of LOFAR. An additional zoom out panel shows GOES X-ray flux for several hours either side of the burst in energy channels 0.5-4.0~$\AA$ (blue) and 1.0-8.0~$\AA$ (red).}
\label{fig:lofar_psp_typeIII}
\end{figure}

\begin{figure}[t!]
\includegraphics[width=0.5\textwidth]{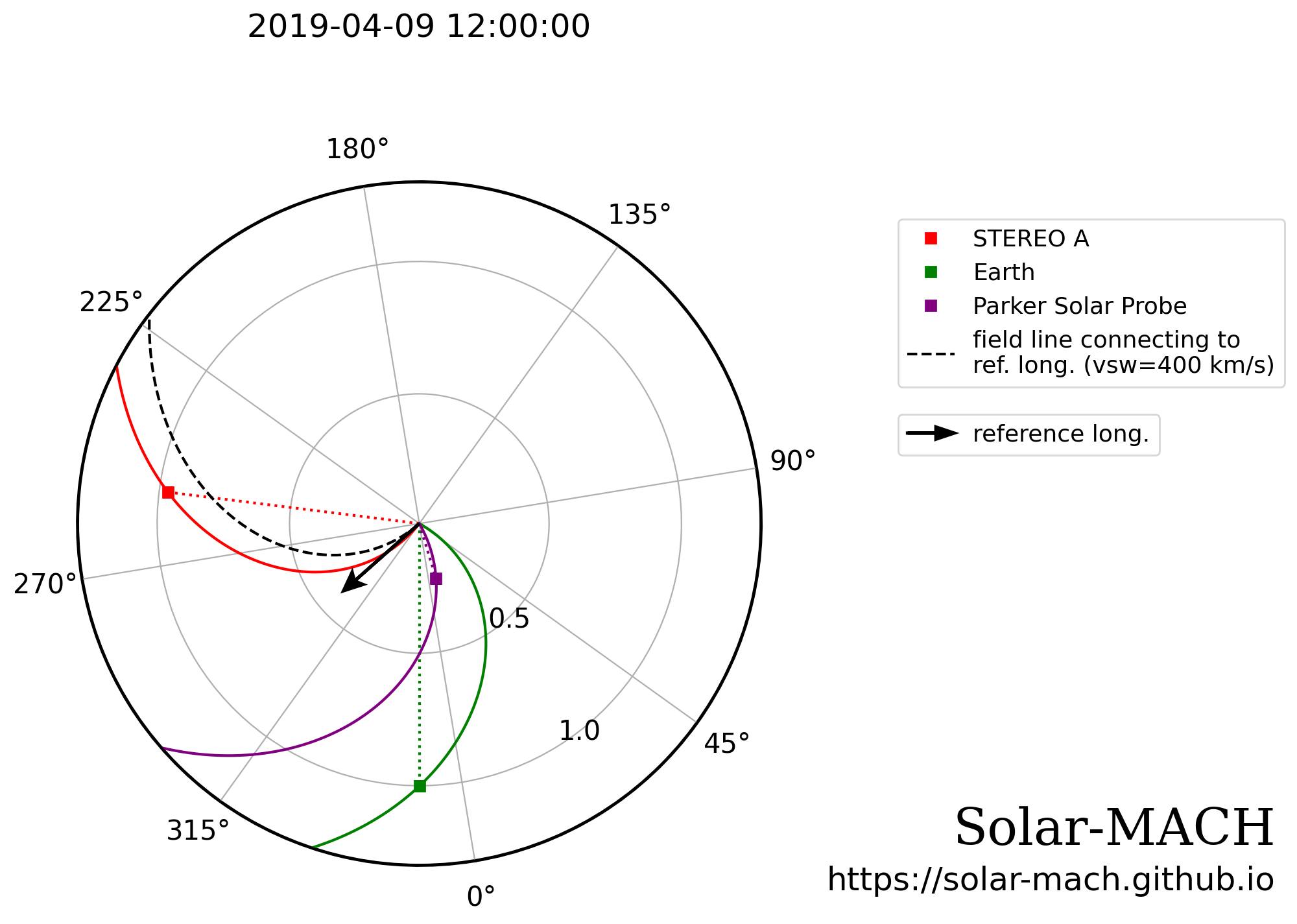}% This is a *.eps file
\caption{Instantaneous ecliptic positions of the spacecraft constellation used in this work at 12:00~UT on 2019/04/09, expressed in Stonyhurst coordinates (Sun-Earth line at 0$^o$ longitude). Each spacecraft is connected back to the Sun by a 400 km/s reference Parker spiral. A black arrow and black dashed Parker spiral indicates the longitude and nominal trajectory of emission from AR 12738. Note the ``Earth'' location is a proxy for the locations of the Wind/WAVES, GOES-15/XRS and LOFAR instruments. Figure generated using \url{https://solar-mach.github.io}}
\label{fig:solar-mach}
\end{figure}

On 2019 April 9 the NOAA AR 12738 had an alpha ($\alpha$) configuration of its photospheric magnetic field containing only unipolar sunspots. It was located east of the central meridian and near the solar equator as viewed from Earth (N06$^o$E48$^o$ at $\sim$18:00 UT and high magnetic flux density plage extending at least 10$^o$ further eastwards, see Figure~\ref{fig:hmi}). At this time AR 12737 was near the west limb and decaying rapidly, consisting only of a magnetic plage. The soft X-ray 1-8 \AA$\,$ flux measured by the GOES X-ray sensor \citep[XRS,][]{Chamberlin2009} showed very small levels of activity, with only B-class flares observable within the time range of our event of interest. The radio and X-ray observations are depicted in  Figure~\ref{fig:lofar_psp_typeIII} with panels (a-d) respectively showing radio spectrograms observed by Wind/WAVES \citep{Bougeret1995}, STEREO A/WAVES \citep[S/WAVES,][]{Bougeret2008}, PSP/FIELDS/RFS and the Low Frequency Array \citep[LOFAR,][]{vanHaarlem2013}. The bottom panel zooms out to a longer time interval and shows GOES X-ray fluxes which are largely featureless both at the time of the burst and in the preceding and following hours. At 12:45\,UT a type III radio burst was observed in the dynamic spectra from LOFAR at 20-90\,MHz, and from STEREO-A/WAVES  and WIND/WAVES  from 0.4-14\,MHz. The same type III was observed by the PSP/FIELDS/RFS instrument 6 minutes earlier, due to its closer proximity to the sun during the event (timeshift applied in the plot, Figure~\ref{fig:lofar_psp_typeIII}c). The relative positions and nominal magnetic connectivity  of the different spacecraft relative to the reference longitude of the source active region is indicated in Figure \ref{fig:solar-mach}. Comparing the radio and X-ray data, this radio burst was not associated with any significant flaring activity. Specifically, no impulsive enhancements above B-class (which was also the background flux level) were observed by GOES at the time of the burst. The burst was therefore likely related to an ongoing noise storm and type III storm associated with AR 12738.

LOFAR observations of the type III from 20-80\,MHZ measure the burst's radiation as it emerges in the lower corona, allowing it to be tracked out to $\sim$3 R${_\odot}$ (section \ref{subsec:lofar}). The observations of the same type III radio burst by PSP, STEREO-A and WIND provide the opportunity to localise the position of this radio burst further out into the inner heliosphere. \textit{In-situ} observations of energetic electrons by the STEREO A Solar Electron and Proton Telescope \citep[SEPT; ][]{MullerMellin2008} and lower energy electrons by the STEREO-A Suprathermal Electron instrument \citep[STE; ][]{Lin2008} suggest the burst electrons from this type III burst can be tracked from its origin, low in the corona at AR 12738, far into the heliosphere all the way to STEREO-A (see Figure \ref{fig:solar-mach}). An energetic electron enhancement of solar origin was indeed detected at STEREO A at compatible times, although as discussed in section \ref{sec:discussion}, we do not conclusively isolate the individual contribution of this burst from the overall population produced by the ongoing type III storm.

\section{Methods and Results} 
\label{sec:methods}

In this section we present the methods employed to track the type III burst from the corona to 1~au and the initial results of our analysis.

\subsection{Imaging and deprojection of LOFAR radio sources}
\label{subsec:lofar}

\begin{figure}
\includegraphics[scale=0.45, trim=1cm 1cm 0cm 0cm]{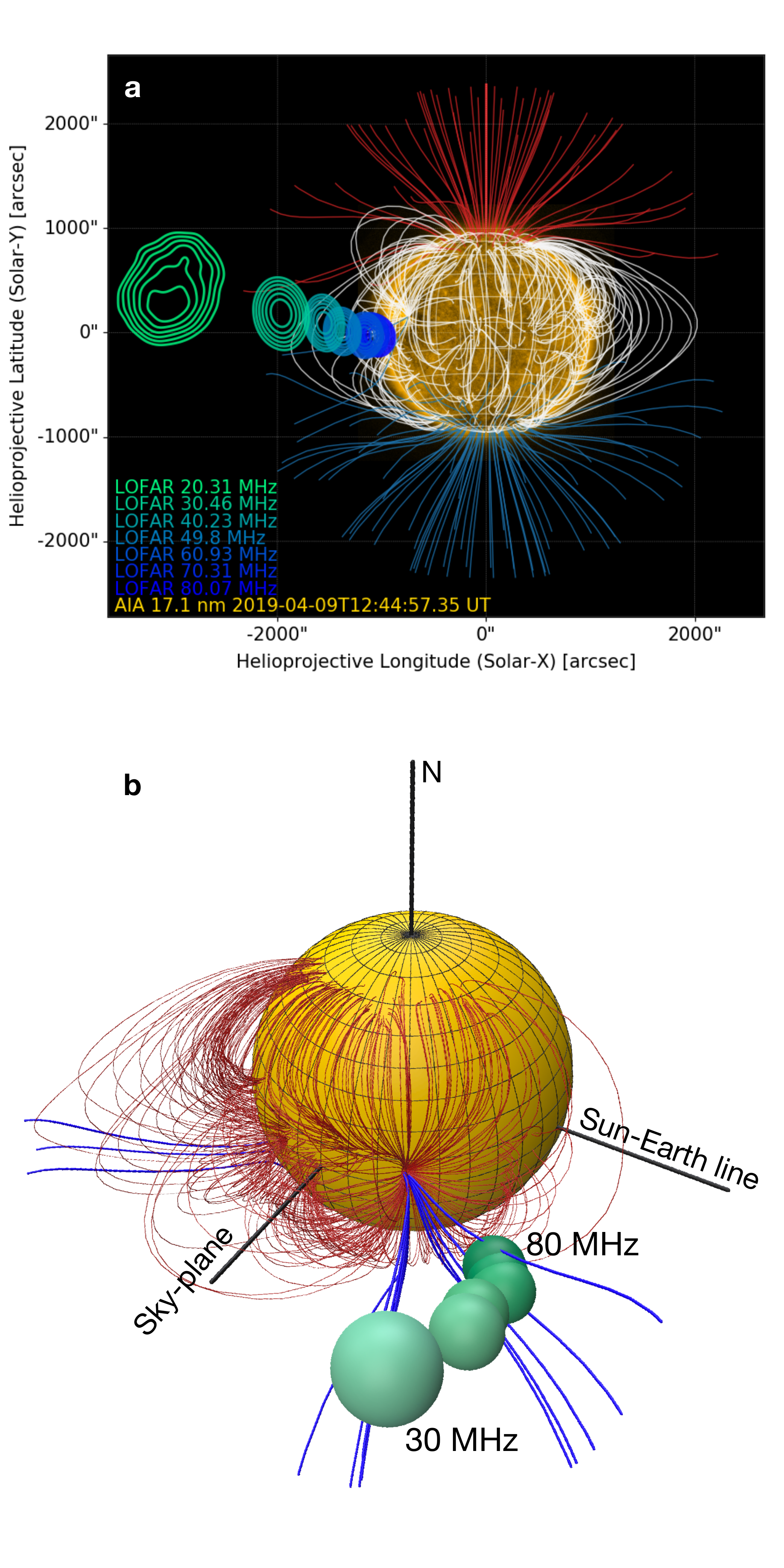}
\caption{(a) Type III radio burst sources observed by LOFAR from 20-80\,MHz overlaid on a PFSS magnetic field extrapolation using GONG magnetograms. The type III sources emanate from the active region near the east limb and are situated close to the negative open magnetic field rooted in the active region. (b) A deprojection of the radio sources using a Newkirk density model \citep{Newkirk1961}, assuming emission at the second harmonic of the local plasma frequency. This places the radio sources on the open field lines $\sim$40$^{\circ}$ which appear to originate from AR12738. This agrees well with the trajectory of the radio sources in Figure~\ref{fig:psp_sta_wind}. In panel (b) field line tracing is performed with a higher density of grid points over a smaller area, hence more open field lines are resolved.
}
\label{fig:lofar3d}
\end{figure}

During PSP encounter 2 the 36 core + remote stations of the LOFAR-provided interferometric observations of the sun, generating imaging spectroscopy observations in the range of 20-80\,MHz (the remainder of the international array was also used to perform beam-formed observations of interplanetary scintillation during this observing period). A dynamic spectrum of the type III radio burst observed by LOFAR is shown in Figure~\ref{fig:lofar_psp_typeIII}(d). LOFAR interferometric data was calibrated (from simultaneous observations of Tau A) and processed using the Default Preprocessing Pipeline \citep[DPPP;][]{vandiepen2018} followed by an implementation of w-stacking clean \citep[WSClean; ][]{Offringa2014} to produce images with a cadence of 1.67\,seconds. Images at seven frequencies from $\sim$20-80\,MHz are displayed in Figure~\ref{fig:lofar3d}a, showing radio sources overplotted on an AIA 171\,\AA\, image and a potential field source surface (PFSS) magnetic field extrapolation using pfsspy \citep{stansby2020} and data from the Global Oscillation Network Group project \citep[GONG;][]{Harvey1996}. The time of the GONG magnetogram used for the extrapolation was 13:04\,UT, on April 9th 2019, with a  source surface radius of 2.5\,R$_{\odot}$. The type III radio sources clearly emanate from AR 12738 along the negative open field to the south of the active region.

In order to determine the location of the type III radio burst sources with respect to the parent active region, a basic deprojection using the Newkirk density model \citep{Newkirk1961} was performed. We assume the type III emission is from harmonic emission in an atmosphere described by a coronal density model where the sources are located at a height of $r_{model}$. Their distance from the plane of sky (POS) is then $z=\sqrt{r_{model}^2-r_{pos}^2}$, where $r_{pos}$ is the height of the radio sources on the POS. The results of this deprojection are illustrated in Figure \ref{fig:lofar3d}b. Although it is a rudimentary deprojection, it places the type III radio burst sources where we would expect them, in an open field region (from a PFSS model) accessible to an active region. The same procedure was performed using the 3.5$\times$Saito density model \citep{saito1977}, which is appropriate for active region radial density profiles \citep{magdalenic2010, magdalenic2012, jebaraj2020}, and the results were the same i.e., the radio sources were deprojected to be within the open magnetic field from AR12738. We also note that the non-radial propagation of the type III sources across the field lines in Figure \ref{fig:lofar3d} may not be physical. This is due to the various uncertainties in source positions on the sky-plane and in the deprojection method which of course neglects all non-radial density inhomogeneities.

The angle from the sky-plane of the type III burst roughly matches where the interplanetary type III is located from PSP-SWAVES-WAVES time-of-flight analysis (section \ref{subsec:TDOA}, below). A few caveats are noted however: (i) the deprojection uses an arbitrary, and spherically symmetric, density model (one of many possibilities), so gives only a rough indication of the source positions, (ii) the radio sources are also large at these frequencies, the sizes of the green spheres in the 3D plot are 1-sigma from a Gaussian fit to the sources in the LOFAR images. 

\subsection{Time-of-flight using PSP, STEREO and Wind}
\label{subsec:TDOA}

As noted above, this burst was also observed at frequencies below $\sim$14~MHz by the three space-based radio instruments S/WAVES, Wind/WAVES and PSP/FIELDS (see Figure \ref{fig:lofar_psp_typeIII}). These radio waves do not penetrate the Earth's ionosphere so are not visible to ground based instrumentation with interferometric imaging capabilities, and such capabilities are not yet available in space. Instead, to continue tracking the bursts we need to employ analysis methods on the more coarsely timestamped dynamic spectra produced by our current space based assets.

In this work, we make use of the fact that the burst is travelling over interplanetary distances and thus the length scales of the trajectory and variation in light travel time to  the different observers are on the order of light minutes. Further, all the above instruments produce dynamic spectrum with a cadence of $\lesssim$1 minute (60s for Wind/WAVES, $\sim$38s for S/WAVES and $\sim$7s for PSP), and on the date of this case study were all located multiple light minutes apart. Therefore, the difference in light travel time from the burst to the different spacecraft is measurable. With the assumption of straight line propagation from a unique source position to each spacecraft, the difference in time of arrival at a given frequency between different spacecraft can be used to make a simple geometric reconstruction of the most likely source position \citep[see e.g.][and Appendix \ref{appendix:tdoa}]{Steinberg1984,Reiner2009,Alcock2018}. With this method and a constellation of three well-separated instruments, the radio burst position may only be confined uniquely in 2D and so we make the assumption that the source is approximately located in the ecliptic plane. This assumption not generally valid but is supported for this specific burst by  (1) the near-equatorial parent active region, (2) the associated coronal open field lines and (3) the imaging results of LOFAR for this burst (Figure \ref{fig:lofar3d}) which place the burst around $\sim4^o$ of the ecliptic plane.

The result of such a reconstruction as a function of frequency for the burst studied in this paper is depicted in Figure \ref{fig:psp_sta_wind}. The implementation procedure is detailed in appendix \ref{appendix:tdoa}. The top panel of Figure \ref{fig:psp_sta_wind} provides an overview of the whole inner heliosphere, depicting the spacecraft constellation triangle consisting of PSP, Wind and STEREO-A, as well as the burst reconstruction coloured from light to dark as it drops in frequency and moves out from the sun. The solid markers indicate the positions corresponding to the central value of the measured time delays, while the diffuse cloud indicates the uncertainty in position obtained by varying the measured time delay within the time resolution of the instruments (see the appendix for more details). Two dotted lines indicate Parker spirals initiated at 30$^o$E and 60$^o$E relative to the central meridian with solar wind speed of 400 km s$^{-1}$, informed by \textit{in situ} measurements at STEREO-A at this time. %These show that it is plausible that energetic electrons injected in the vicinity of these interplanetary field lines would reach STEREO-A (Section \ref{subsec:insitu}). 

The bottom panel of Figure \ref{fig:psp_sta_wind} zooms in on the reconstructed type III trajectory and uncertainty region. Red dots close to the origin indicate the source positions as reconstructed by LOFAR (Section \ref{subsec:lofar}) and plotted in the ecliptic plane. 

It is worth noting the time delay generated trajectory implies the burst has been tracked out to 40 $R_\odot$. However, the lowest frequency observed at PSP is only 800 kHz for which typical density models \citep[e.g. ][]{Leblanc1998} would suggest the outer radius at which emitted radiation reaches PSP is only $\sim$10 $R_\odot$ (20 $R_\odot$) for fundamental (harmonic) radiation.

We note that figure \ref{fig:lofar_psp_typeIII} shows clearly that the burst continues to propagate beyond the maximal triangulated distance, since radio emission is observed in Wind and STEREO-A at frequencies below the lowest frequencies observed by PSP. This suggests the beam reaches regions where the local plasma frequency is below the in situ plasma frequency at PSP, so radio emission cannot propagate inwards towards PSP. Thus, although we are unable to track the burst further with TDOA, it is plausible that electron beams along these mapped field lines can make it out to reach STEREO-A at 1 au.

\begin{figure}[t!]
\includegraphics[scale=0.38, trim=0cm 0cm 0cm 0cm]{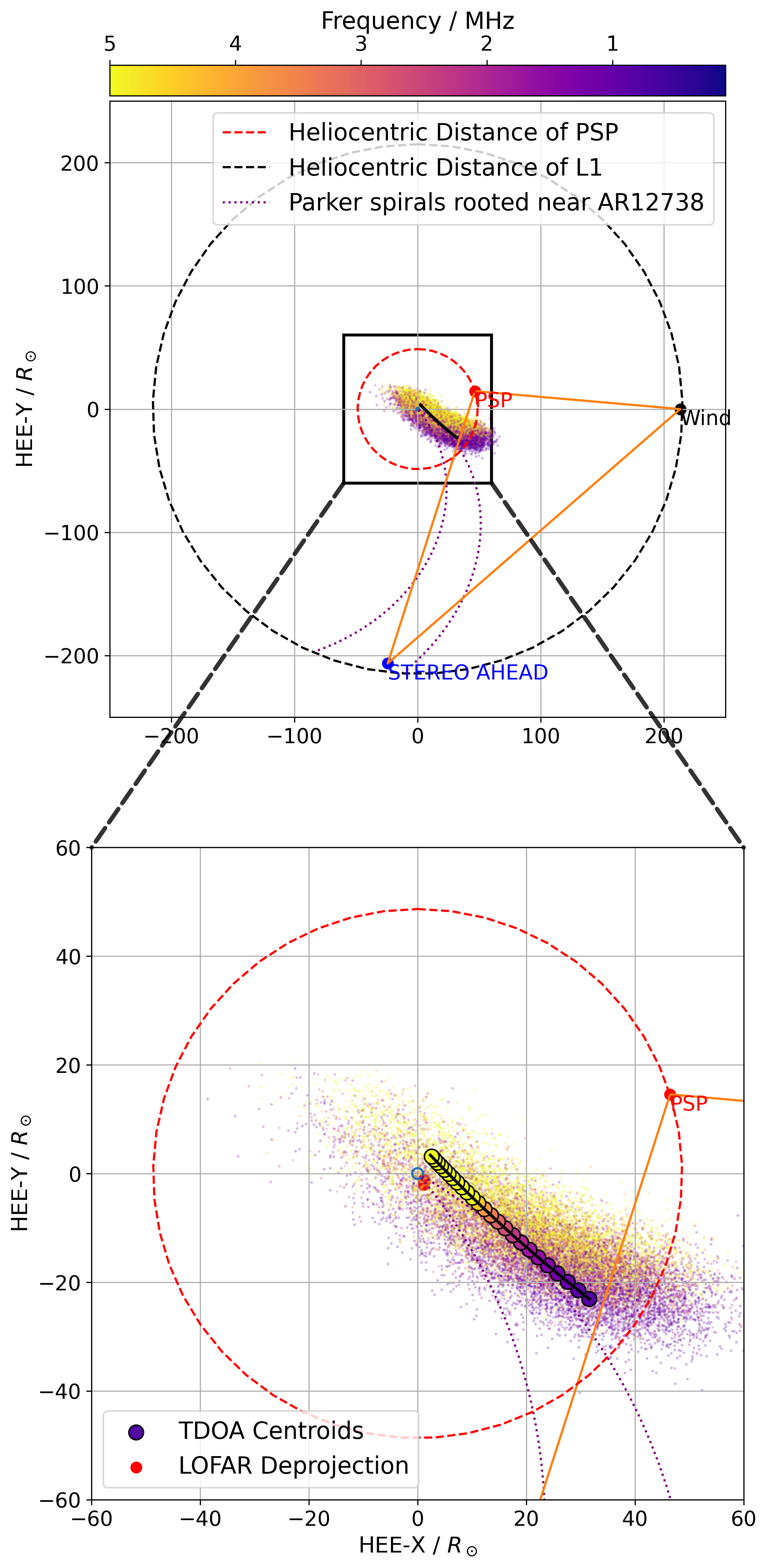}
\caption{TDOA localization in the ecliptic plane of a type III radio burst trajectory in the inner heliosphere as measured by PSP, WIND and STEREO-A. The solid markers in the bottom panel represent the central value, while the diffuse cloud shows the position uncertainty due to the impact of the instrument time resolution on the time-of-flight measurement, both colored by measured frequency. The LOFAR deprojections are visible for comparison in the zoomed in bottom panel. The coordinate frame is Heliocentric-Earth-Ecliptic (HEE).}
\label{fig:psp_sta_wind}
\end{figure}

\subsection{In-situ analysis}
\label{subsec:insitu}

\begin{figure}[t!]
\includegraphics[scale=0.5, trim=2cm 0.5cm 0cm 0cm]{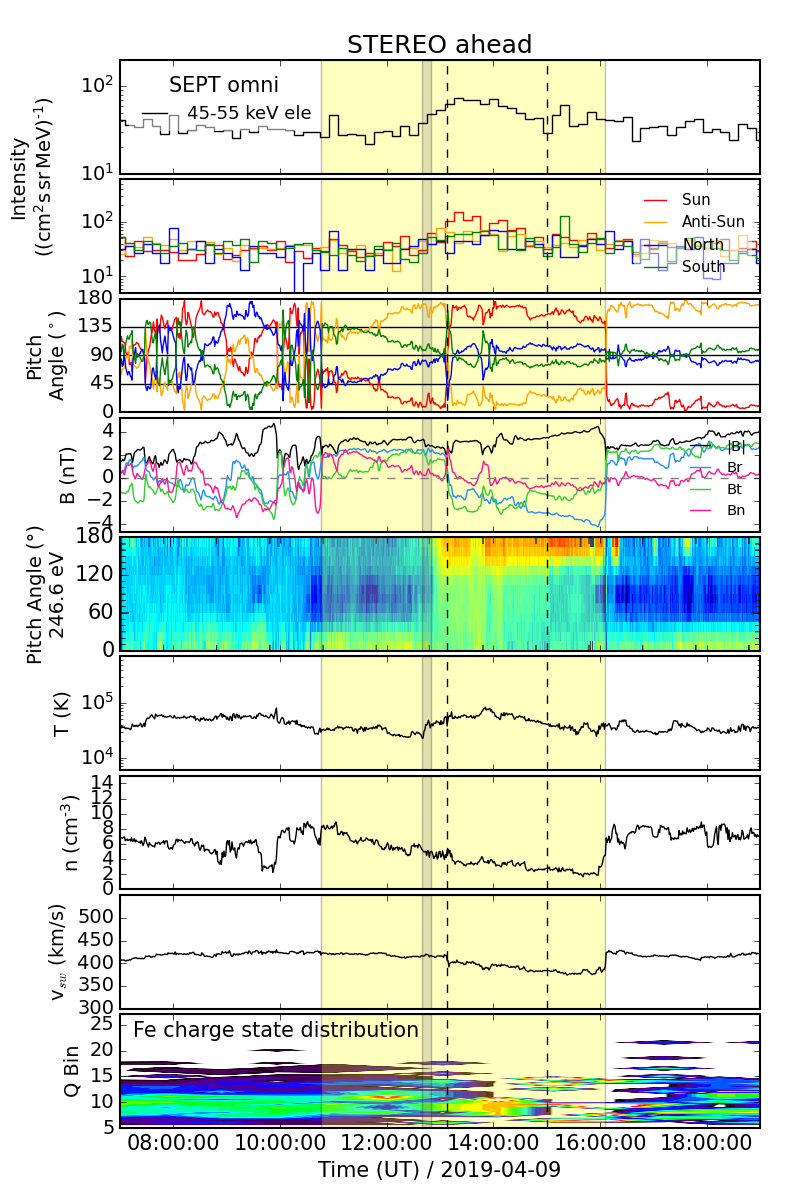}
\caption{\textit{In situ} observations by STEREO-A on 2019 April 9. From top to bottom: Omni-directional 45-55 keV electron intensities, sectored intensities measured by the four telescopes of the STEREO/SEPT instrument (both using 10-minute averaging, respectively). Pitch angles scanned by the central axes of these four-telescopes. Magnitude and components of the magnetic field vector in the spacecraft-centered RTN coordinate system. Pitch angle distribution of suprathermal electrons at 246.6 eV (color for the logarithm of phase space density). Solar wind proton temperature, density, and speed. Distribution of iron charge states $Q_{Fe}$ as measured by STEREO-A/PLASTIC. The yellow shaded period indicates a period with peculiar field direction studied in detail. Grey shaded bar and dashed lines respectively mark the onset of the SEP event (12:40-12:50 UT), magnetic sector boundary crossing (13:08 UT), and return of electron flux to background levels (15:00 UT).}
\label{fig:stereo_sept}
\end{figure}

As mentioned in the previous section, figure \ref{fig:psp_sta_wind} shows that the magnetic connection along nearby Parker spiral field lines would allow the potential \textit{in situ} detection by STEREO-A of the electrons produced during the solar eruption origin of the type III emission, although the exact nominal 400~km/s connection of STEREO A would be westward of the AR (see Figure \ref{fig:solar-mach}).

We have analyzed energetic electrons measurements from the SEPT and the Solar Wind Electron Analyzer \citep[SWEA; ][]{Sauvaud2008} instruments on board STEREO-A to determine whether such electrons were detected. We further contextualize the \textit{in situ} environment around the spacecraft at this time with magnetic field measurements from STEREO-A/IMPACT/MAG \citep{Acuna2008}, thermal plasma measurements from SWEA and composition data with STEREO-A/PLASTIC \citep{Galvin2008}. 

SEPT consists of two dual double-ended particle telescopes which measure electrons in the energy range of 40–400 keV. The four apertures of SEPT were termed Sun, Anti-Sun, North, and South owing to their looking directions along the nominal Parker spiral direction in the ecliptic plane (Sun and Anti-Sun) and perpendicular to the ecliptic plane (North and South).
In the original configuration of the STEREO-A spacecraft, the longitudes of the central axis of the Sun and Anti-Sun apertures in the spacecraft-centered Radial-Tangential-Normal (RTN) coordinate system were $\phi_{Sun}$$\sim$$135^{\circ}$ and  $\phi_{Asun}$$\sim$$-45^{\circ}$, respectively, designed to align with the nominal Parker Spiral azimuthal angle at 1~au.
However, after the STEREO-A superior conjunction with Earth in early 2015, the spacecraft changed orientation resulting in the Sun and Anti-Sun apertures still pointing in the ecliptic plane but now \textit{perpendicular} to the nominal Parker spiral direction, with $\phi_{Sun}$$\sim$$-135^{\circ}$ and  $\phi_{Asun}$$\sim$$45^{\circ}$, respectively. 

On 2019 April 9, a small electron intensity enhancement at energy channels below $\sim$100 keV was observed by STEREO-A/SEPT close to the emission time of the Type III burst studied in the previous section.
Figure~\ref{fig:stereo_sept} summarizes the \textit{in situ} context at the location of STEREO-A at the relevant times. In the top panel, we see timeseries of 45-55 keV electron omni-directional intensities, and in the second panel those of the sectored intensities measured by the different apertures of SEPT. This energy channel was chosen as it most clearly demonstrates the event onset, peak and decay. The onset of the electron event is most clearly seen in the omni-directional intensities (top panel). We identify the onset to be between 12:40~UT and 12:50~UT, marked by the grey shaded range which reflects the coarse time resolution of the SEPT instrument. The current sheet crossing (sector boundary) indicated by the magnetic field orientation changing to $\phi\sim-135^{\circ}$ occurred at $\sim$13:08~UT (first) dashed line) when electron intensities were still increasing. Electron intensities reached a maximum at about $\sim$13:30~UT principly in the `Sun' aperture (red trace), and slowly decreased to pre-event intensities at about $\sim$15:00~UT (second dashed line). 

The third panel of Figure~\ref{fig:stereo_sept} shows the angle between the instantaneous magnetic field direction (RTN components shown in the fourth panel of the figure) and the central axis of each one of the SEPT apertures. These panels show that throughout all this period, the magnetic field lay close to the ecliptic and perpendicular to the nominal Parker spiral field direction, which makes the post-superior-conjunction orientation of the Sun and Anti-Sun apertures of SEPT ideal for the detection of these electrons (as indicated by the orange and red trace's close alignments with $0^\circ$ and $180^\circ$ in the pitch-angle panel). Prior to the current sheet, the ecliptic component of the field was oriented at $45^o$ in the $RTN$ coordinate system, and then flipped $180^o$ to point at $-135^o$. The entire stream containing this ``perpendicular-to-Parker'' interval is captured by the yellow box superimposed on the different panels.  The field gradually rotates into the ``perpendicular-to-parker configuration from the start of the yellow box through to the electron onset (grey shaded bar). The current sheet crossing happens approximately halfway through this interval. 

The fifth panel down presents the pitch angle distribution of suprathermal electrons at 246.6 eV as measured by SWEA.
Prior to the current sheet crossing, the electron pitch angle distribution (defined relative to the local field orientation) is anisotropic and peaked at $0^o$ indicating it is parallel to the field lines. After the current sheet crossing, the distribution is peaked at $\sim$180$^{\circ}$ i.e. antiparallel to the field lines. Given the field orientation has flipped $180^o$, this indicates the dominant direction of the suprathermals remains consistent and at all times is flowing into the direction of the SEPT ``Sun'' aperture. After the current sheet and coincident with the peak in 45-55~keV electrons, the suprathermal electrons do become more isotropic (but still clearly single-peaked); this pitch angle spread decays away on the same timescale as the energetic electron population decreases to background levels. We note the pitch angle distributions of the 246.6~eV electrons are consistent with the arrival direction of the energetic electrons (flowing into the SEPT ``Sun'' aperture).

Plasma temperature, density radial flow speed are shown respectively in the next three panels of Figure \ref{fig:stereo_sept}. There are no sharp changes in these quantities during the energetic electron enhancement. A small but steady rarefaction in solar wind speed occurs across the enhancement. The last panel shows the iron charge state distribution ($Q_{Fe}$) measured by PLASTIC. During the electron enhancement the the total flux across different iron charge states increased and the distribution was specifically concentrated around charge states of of 9-15. This contrasts with ambient solar wind prior to and following the event where the distribution is flatter and at a lower overall flux.

To relate the \textit{in situ} measurements to the radio burst timing at high frequencies, it is important to consider the interplanetary travel times of the burst electron beams. A 45 keV (55 keV) electron would take about $\sim$25 ($\sim$ 23 minutes) to travel a  distance of 1.2 au along a nominal Parker spiral for a spacecraft located at 1 au from the Sun.  For an observer located at 0.967 au like STEREO-A, a 45 keV (55 keV) electron would take about $\sim$20 ($\sim$ 19 minutes) to travel  the radial distance from the Sun to the spacecraft. The actual path followed by these electrons before reaching STEREO-A when embedded in this structure, as well as how this structure modified the STEREO-A magnetic connection with the Sun, are uncertain. This means mapping arrival time of the event at STEREO-A back to a specific emission from the Sun is difficult. However, since the nominal Parker spiral  represents the simplest possible field configuration, the arrival times quoted above are likely to represent a lower bound on expected time delays.

The first radio waves associated with the type III radio burst were detected by LOFAR at 12:45~UT implying a release time in the low corona of $\sim$12:37~UT. This appears too late therefore to explain the initial onset time of the energetic electrons observed in situ by STEREO-A/SEPT at $\sim$12:40-12.50~UT. Moreover, the timescale of the whole enhancement tracked by SEPT is of order 1-2 hours, significantly longer than the timescale of any given burst. 
However, the anisotropic electron enhancement which continued during the current sheet passage at 13:08~UT and the subsequent electron intensity peak at $\sim$13:30~UT are compatible with electrons injected during the type III radio burst analyzed in Section~\ref{sec:methods}. This timing issue is discussed further in the following section. 

\section{Discussion}
\label{sec:discussion}

In this work we have presented a multi-instrument analysis where we tracked a single well-isolated, near-ecliptic-propagating type III radio burst. Specifically, we followed its injection onto open field lines by a source active region, its propagation through the corona and out into interplanetary space and lastly, its possible association with \textit{in situ} measurements from the STEREO-A spacecraft near 1~au. In the following subsections, we discuss the implications and our interpretation for each part of the analysis.

\subsection{Emergence from the Corona}

Using radio imaging with the Low Frequency Array from $20-80$ MHz and a simple deprojection analysis assuming a Newkirk coronal electron density model \citep{Newkirk1961}, the burst was placed in 3D space at coronal altitudes at longitudes clearly associated with AR 12738 located at N06$^o$E48$^o$ in helioprojective coordinates on April 9, 2019 (Figure \ref{fig:hmi}). This was consistent with the association of the overall type III activity with this AR on these dates made by \citet{Pulupa2020,Krupar2020,Cattell2021}. The imaging showed the burst increasing in altitude with decreasing frequency and the deprojection placed it in a region of open magnetic field lines in the vicinity of the relevant active region (figure \ref{fig:lofar3d}). The deprojection analysis was repeated with the use of the 3.5xSaito density model \citep{saito1977} and the association of the burst with AR 12738 and extrapolated open field lines persisted along with the approximate coronal source altitudes and overall directivity. However the limitation of a spherically symmetric density profile leaves an unquantified error in the plane of sky distance. 

GOES/XRS measurements of x-ray radiation over this time indicated that there was no significant flaring component associated with the injection of the burst (figure \ref{fig:lofar_psp_typeIII}). This suggested non-flaring processes such as flux emergence and small-scale reconnection are responsible for the studied type III. This case study therefore represents an example of non-flare-associated interplanetary type III burst.

Additionally, many faint type III bursts in LOFAR preceding and following this burst are not observed at interplanetary frequencies although some are measured only by PSP between 1-10~MHz (see panel c) and not by Wind or STEREO A. This suggests that more bursts are being injected onto interplanetary field lines than are visible in radio at 1~au during this time interval, and again without flare signatures. As discussed below, timing analysis of the \textit{in situ} data at STEREO-A shows energetic electrons arrived earlier than can be explained by the tracked burst timing; earlier injections which are not bright in radio are a possible resolution to this.

\subsection{Interplanetary Propagation}

Having established the burst to be injected onto open field lines and therefore expected to reach interplanetary space, we investigated the source propagation out to several 10s of solar radii using a novel time delay technique termed Time Delay of Arrival  (TDOA; see appendix \ref{appendix:tdoa}) leveraging the multi-light minute separation between the different heliospheric spacecraft with radio spectrograph instrumentation (PSP, Wind and STEREO-A at the time of the studied event). The method assumed that the burst was located in the ecliptic plane along with the observing spacecraft, and that the radio waves free streamed from source to observer. These sources of uncertainty are discussed further in appendix \ref{appendix:tdoa} where we conclude the former is well justified for this burst, while the second is harder to quantify.

This analysis yielded a trajectory for the burst at frequencies from 10 MHz down to 800 kHz (figure \ref{fig:psp_sta_wind}) at which point the radio frequency became evanescent along paths to PSP (this means the burst propagated beyond the heliocentric distance of PSP). The burst was still observed at lower frequencies by the 1~au spacecraft which shows that it continued propagating futher out into space, and is therefore plausible it would survive out to STEREO-A.

The derived trajectory showed the burst to be increasing in distance from the Sun with decreasing frequency, and to be directed along a longitude consistent with the active region and coronal trajectory found by LOFAR, as well as Parker spiral field lines connecting to the vicinity of STEREO A. As part of making these conclusions we utilize error regions computed by accounting for the limited time resolution of the individual instruments radio data products, most significantly that of Wind/WAVES at 60s (see appendix \ref{appendix:tdoa}). This error region is large and extended in the radial direction (see bottom panel, figure \ref{fig:tdoa_schematic}) which leads to the error regions of adjacent frequency channels significantly overlapping. However both the centroid and radial extrema of the error region do shift outwards with decreasing frequency (the latter is not distinguished in figure \ref{fig:psp_sta_wind} as the error regions are shown on a continuous color scale). The azimuthal extent of the error region is smaller but still significant and is similar in scale to the apparent longitudinal offset between the TDOA centroids and the deprojected longitudes from the LOFAR imaging results. However, the small misalignment is also possibly contributed to by innacuracies in the LOFAR deprojection analysis.

The heliocentric distances computed from the TDOA analysis were significantly larger than expected from typical interplanetary density models. A potential explanation for this is propagation effects such as scattering by density fluctuations at the emission location (i.e. a breakdown of the above assumption of free streaming). This mechanism has been shown to make coronal radio sources appear to be lensed out to higher altitudes \citep{Kontar2019}. \citet{Krupar2018} and \citet{Krupar2020} have shown that density fluctuations also scatter type III radio emission at interplanetary frequencies and in particular play a key a role in shaping the exponential tail of type III flux-time profiles (see e.g. figure \ref{fig:time-profile}). To further investigate this lensing interpretation, the TDOA method presented here should be applied to a large statistical sample of ecliptic-propagating type III bursts in future work.

We conclude that the TDOA method as applied here provides real and useful information about the propagation of type III bursts on interplanetary distances. A necessary condition of its use was justifying the strong assumption of ecliptic propagation of the burst. It was also required that the difference in light travel time to the different spacecraft observing the burst was at least several minutes, such that differences in arrival time could be robustly resolved. With these conditions satisfied, TDOA allowed association of the burst with an approximate source region, interplanetary propagation direction, and an estimate of its radial motion. As discussed quantitatively in Appendix \ref{appendix:tdoa} large uncertainties from instrument resolution appear to be the dominant error source, although the impact of propagation effects was not quantified. Nevertheless, the method is appealing since the construction does not make any assumption about density models or fundamental vs. harmonic emission, although the assumption of free streaming propagation needs further investigation and quantification.

\subsection{Signature at 1 au}

\begin{figure}[t!]
\centering
\includegraphics[width=0.5\textwidth]{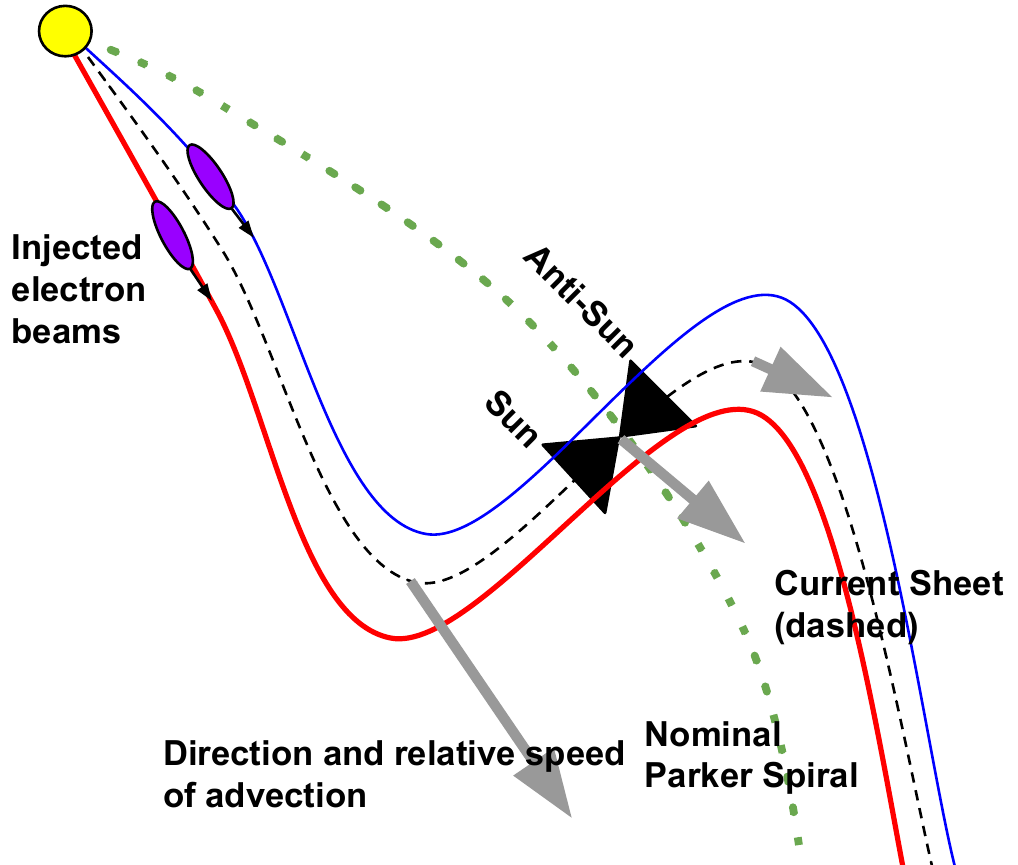}
\caption{Proposed schematic of stream structure at STEREO A during the time interval studied (yellow shading in figure \ref{fig:stereo_sept}). STEREO A/SEPT FOVs indicated by the 2 black triangles. Red and blue kinked field lines of opposing polarities. Nominal Parker spiral indicated by green dotted curve . Purple ellipses with arrows indicate electron beams injected onto these field lines. Grey solid arrows indicate the proposed flow velocity across the kink and the overall advection direction of the structure.}
\label{fig:schematic}
\end{figure}

Lastly, motivated by the consistent longitude between the active region, LOFAR imaging, interplanetary triangulation, and Parker spiral field lines which mapped to the location of STEREO-A, we examined in detail the signatures of energetic electrons at that spacecraft to determine if the electron beam was detected at 1~au (figure \ref{fig:stereo_sept}). An energetic electron event was indeed observed at STEREO-A at compatible times, although its true onset (12:40-12.50~UT) was too early to be explained by the specific burst studied here (emitted at approximately 12:37~UT). The duration of the electron enhancement (peaking at 13:30~UT) is compatible with the electrons that generated the studied type III radio burst arriving at STEREO-A during the rising phase of the energetic electron event and suggests they contribute to the peak of this electron event. 

Further, detailed \textit{in situ} analysis of the solar wind stream that swept past STEREO A at this time demonstrated clear evidence that the energetic electrons measured were solar in origin and were not associated with any \textit{in situ} electron energization processes. Specifically, evidence for solar origin included that 1) the energetic electron flux peaked highest in the sunward detector implying it was flowing from the sunward direction, 2) the electron pitch angle was singularly peaked at all times indicating open magnetic field lines, and 3) the iron charge state variation indicated a distinct solar source for the duration of electron enhancement \citep{Lepri2004}. Evidence that the electrons were not energized in situ included the lack of shock signatures in the solar wind parameters, and the lack of flux ropes, stream interaction regions or CME observations prior to or following the studied event.

Although there is good evidence that the electron enhancement was solar in origin, the magnetic configuration of the stream was unusual. Specifically, the measured magnetic field lines at the event time were perpendicular to the nominal Parker spiral direction and also included a current sheet crossing where the field direction flipped 180 degrees. This orientation was rotated into gradually and reached the ``perpendicular'' direction at the time the electron burst began.

In figure \ref{fig:schematic} we present a schematic representation of such a stream configuration which could explain the \textit{in situ} signatures: Here, we propose a strong kink in an otherwise nominal set of Parker spiral field lines containing a current sheet to explain both the field orientation and the in situ evidence for an open field line connecting back to the sun. The orientation of the STEREO-A SEPT Sun and Anti-Sun apertures are also illustrated as black triangles, along with the direction of electron beam motion which shows how the instrument orientation is fortuitous for making the observations in this picture. As time passes, this kink would advect out radially anti-sunwards and over STEREO A,  causing a current sheet crossing to be measured. The longitudinal mapping of the field line would be affected by the presence of such a kink, with the size of the shift determined by the transverse scale of the kink. With the kink oriented in the sense shown here, the extra shift relative to the nominal Parker spiral is in fact Eastwards, which is the right direction to account for the slight misalignment indicated in Figure \ref{fig:solar-mach}.  Such a kink can be supported by a velocity shear \citep[][]{Schwadron2021}. For the schematic shown in figure \ref{fig:schematic}, the velocity shear is oriented as indicated by the size of the large gray arrows at the inflection points of the curve. This leads to the expectation that STEREO A would measure a gradual decrease in solar wind velocity with time as the kink washes over it which is consistent with the in situ data  (figure \ref{fig:stereo_sept}).

The timing analysis mentioned in section \ref{subsec:insitu} implies the actual electron beam studied in this work would arrive during the rise in electron flux and potentially contributes to the overall peak at 13.30~UT. We note that although the nominal arrival time along a Parker spiral would be around 12.57~UT for a 12.37~UT emission, the perpendicular field configuration combined with the evidence that the field line is open and connects to the sun implies a longer travel time since all field line geometries consistent with these two observations have a longer path length than the nominal spiral.

Taken all together, we interpret these observations as showing that the studied burst contributes a subset of the population of energetic electrons measured by STEREO-A, but that the overall population forming the enhancement (top panel, figure \ref{fig:stereo_sept}) was comprised of multiple type III ejections of varying radio brightness which were happening near continuously \citep[figure \ref{fig:lofar_psp_typeIII}d and ][]{Cattell2021}. In this picture, STEREO A only measures this population when a favorable stream configuration which connects back to the source active region advects over the spacecraft (for example, the configuration proposed in figure \ref{fig:schematic}). Thus the onset of the electron enhancement would not relate to a specific injection time, but rather the time when the ``perp-to-parker'' field configuration was reached and connected STEREO-A to the source active region.

It is interesting to note the energetic electron enhancement shows a smooth enhancement and takes place over a timescale of hours. There are no sudden enhancements in flux which would allow the contributions of individual beams to be identified as they arrived. This provides further evidence that the observations indicate the result of multiple injections smoothed together, which could occur via processes such as perpendicular diffusion \citep{Strauss2017}. This situation contrasts to previous studies where clear linkage was observed. For example, impulsive ($\mathcal{O}($10~min)) electron enhancements have been observed from similar periodicity type IIIs \citep[e.g.,][]{Klassen2011}. In other cases \citet{Krucker1999} and \citet{Haggerty2002} identified near-relativistic electron events for which the estimated release time of the electrons was delayed with respect to the onset of the type III radio bursts, but still found the electron event could be associated with a specific burst after accounting for this delay. Such combinations of delayed emission could also contribute to the merging of arrival times of many electron beams.

\subsection{Summary and outlook}

In closing, we have studied and tracked a non-flare related interplanetary type III burst from its active region source, its injection onto open coronal field lines, its propagation along interplanetary field lines and shown evidence that it contributes to an energetic electron enhancement measured by STEREO-A at compatible times. Contextual in situ data taken by STEREO-A showed that the local magnetic field was oriented perpendicular to the Parker spiral which suggests an off-nominal connectivity. We suggested an interpretation (figure \ref{fig:schematic}) to explain this observation of kinked interplanetary field lines tracking back to the source active region. 

A motivation for developing the interplanetary TDOA method was to use type III bursts as passive tracers of magnetic field lines in an application similar to \citet[][]{Li2016}, and to do so in a density-model independent way. Such tracing could then be applied to differentiate more complex magnetic models than the simple Parker spiral \citep[e.g. ][]{Tasnim2019}. It could also in principle operate as a direct way to constrain non-Parker field geometries and test \textit{in situ} interpretations such as the one made in figure \ref{fig:schematic}. As yet, this goal has not been achieved as evidenced by the much larger than expected heliocentric distance of the source positions. Further work is needed to investigate the uncertainties, most notably the unquantified divergence from radio free streaming.

To bring this story together we needed to make use of complementary observations including solar magnetograms, GOES X-ray observations, LOFAR ground based radio imaging observations, interplanetary radio dynamic spectra from widely separated interplanetary spacecraft and lastly, the powerful suite of \textit{in situ} instruments on board STEREO A. This demonstrates the power of using the heliospheric systems observatory (HSO) for connecting the inner heliosphere and solar corona and demonstrates a useful future combination of techniques for tracking type III burst both for studying their intrinsic properties, as well as the convolved information they carry about the medium they propagate through. It is important to reiterate that near ecliptic propagation of the burst was a strong assumption for the interplanetary portion of this work, and needs to be evidenced to re-apply the 3-spacecraft TDOA method to future events. However since the launch of Solar Orbiter in 2020 \citep[SO; ][]{Muller2020}, there is a 4th source of interplanetary radio observations, leading to the possibility of extending this part of the analysis to full 3D localization  Combined techniques as shown in this paper will therefore continue to become more capable and powerful as our instrumentation of the heliosphere continues to become more complete.

\acknowledgements
Parker Solar Probe was designed, built, and is now operated by the Johns Hopkins Applied Physics Laboratory as part of NASA's Living with a Star (LWS) program (contract NNN06AA01C). The FIELDS experiment on PSP spacecraft were designed and developed under NASA contract NNN06AA01C. This work uses data from GONG, WIND, STEREO, LOFAR and GOES-15. The authors thank the mission and project teams and the Space Physics Data Facility for making the data publicly available. S.T.B was supported by NASA Mary W. Jackson Headquarters under the NASA Earth and Space Science Fellowship Program Grant 80NSSC18K1201. L.K.J thanks the support of the STEREO mission. D.L. acknowledges support from NASA Living With a Star (LWS) programs NNH17ZDA001N-LWS and NNH19ZDA001N-LWS, and the Heliophysics Innovation Fund (HIF) program. 
N.D. is grateful for support by the Turku Collegium for Science, Medicine and Technology of the University of Turku, Finland. E.C. is supported by the Schr\"odinger Fellowship at the Dublin Institute for Advanced Studies.
\appendix

\section{Time Delay of Arrival (TDOA) Technique}\label{appendix:tdoa}

\begin{figure*}
    \centering
    \includegraphics[width=0.7\textwidth]{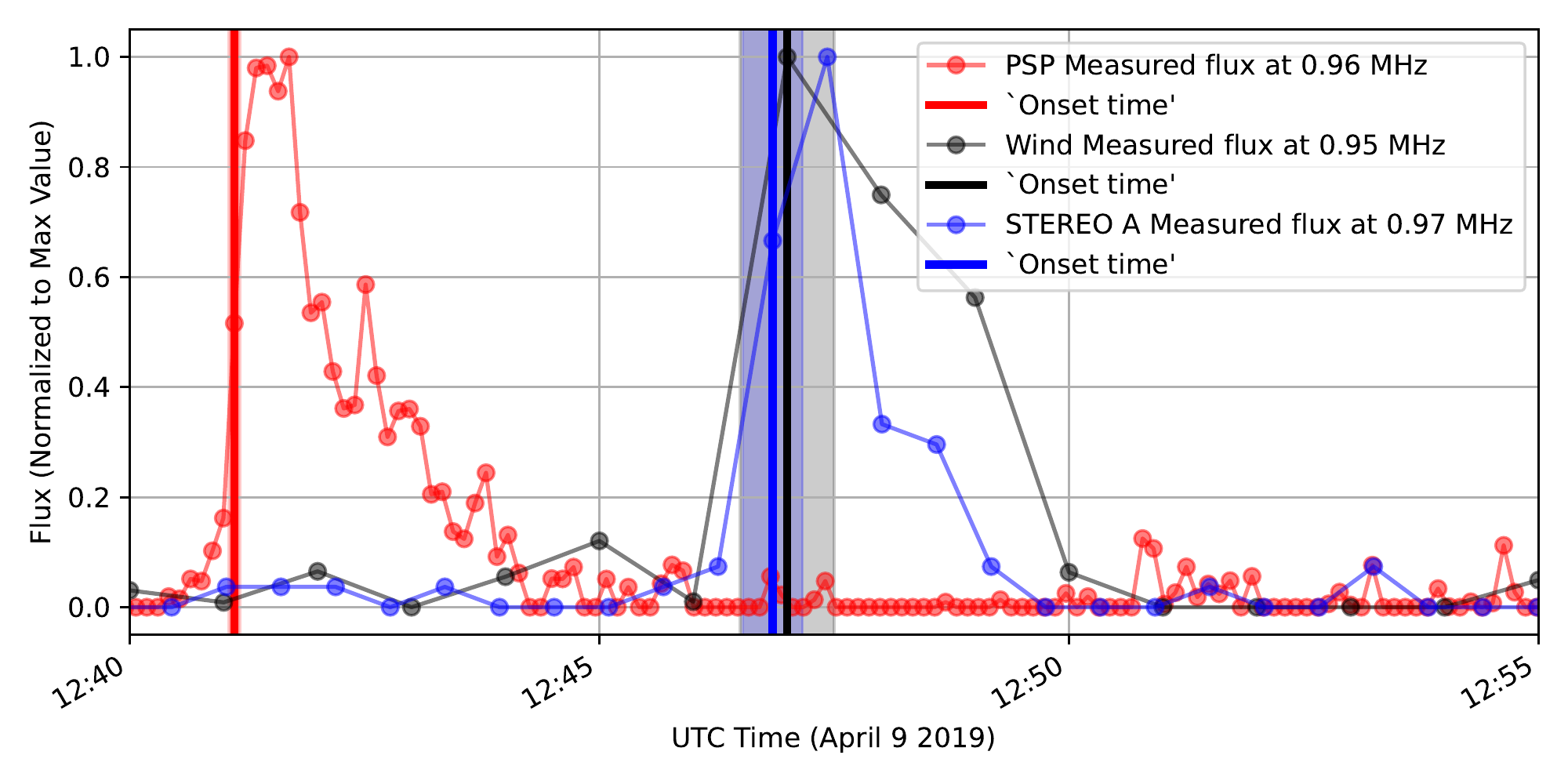}
    \caption{Example radio flux vs time profile and onset extraction example. Each profile is normalized by its maximum value so they can be shown on the same axes. Profile is from the closest frequency channel to 0.96 MHz in each spacecraft receiver. Red, blue and black profiles respectively show PSP/FIELDS/RFS, S/WAVES and Wind/WAVES observations. Data points are shown explicitly to demonstrate differing instrument resolution.}
    \label{fig:time-profile}
\end{figure*}

In this work, we utilized the time delay of arrival \citep[TDOA; ][]{Steinberg1984,Reiner2009,Alcock2018} technique to derive the trajectory of the type III source region at heliospheric distance scales (i.e. large fractions of an au). In this appendix we explain the methodology used to derive these distance estimates.

\begin{figure*}
    \centering
    \includegraphics[width=\textwidth]{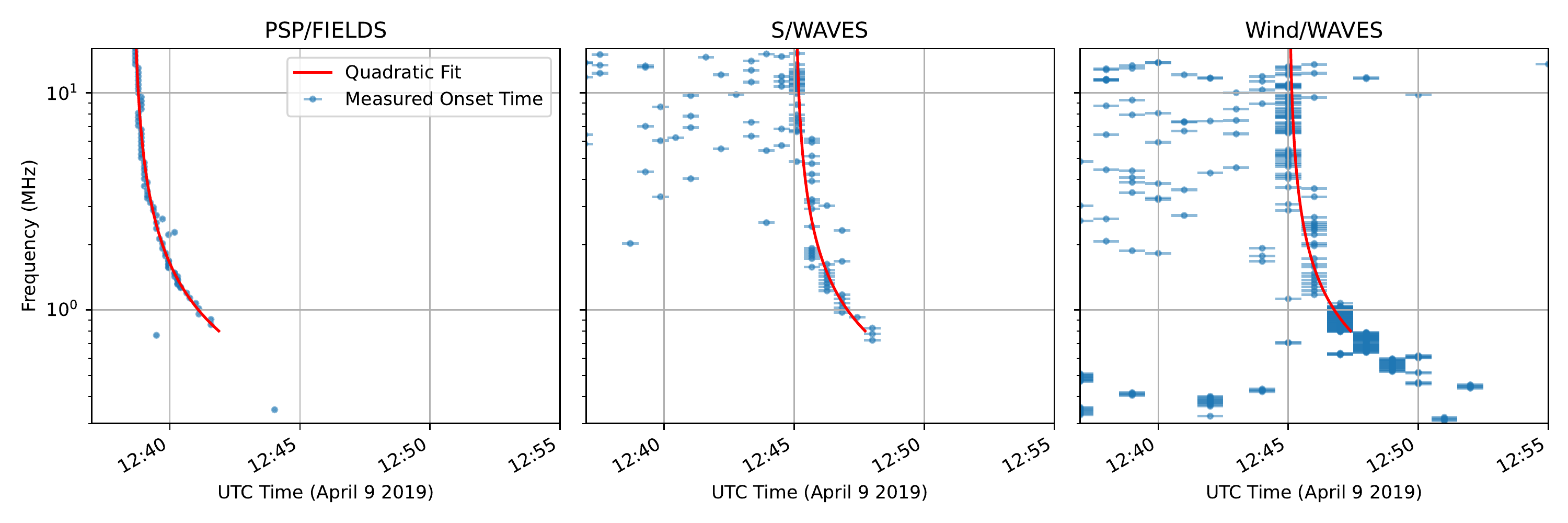}
    \caption{Onset time feature extraction (blue) and quadratic fitting (red) to the type III burst studied in this work as observed by PSP, STEREO A and Wind. Error bars show the instrument time resolution. Fitting is only done over the frequency range clearly observed in all three spacecraft.}
    \label{fig:feature-extraction}
\end{figure*}

\begin{figure}
    \centering
    \includegraphics[width=0.45\textwidth]{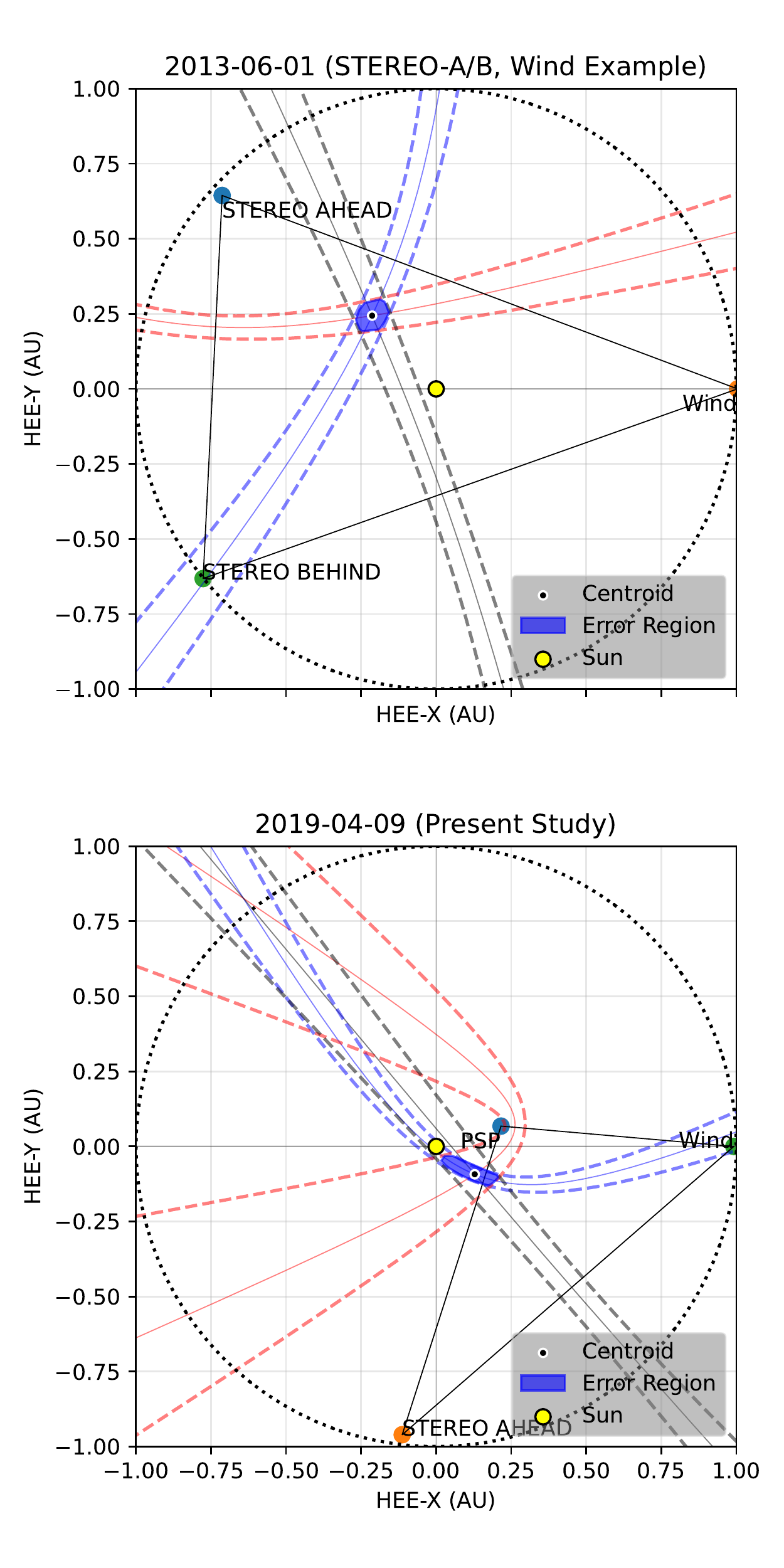}
    \caption{Schematic of the time delay of arrival localization method and associated error region. The top panel shows a synthetic example using the positions of STEREO A/B and Wind in 2013 to clearly show how the hyberbola intersection defines the error region. The right hand panel shows the specific geometry for the event in this study using the measured time delays at 1.04MHz. Each panel shows in the HEE frame: the spacecraft geometry, the hyperbola generated from each pair of time delays, the resulting source centroid and error region.
    Annotations on the plot also include the position of the Sun and a dotted circle indicates 1~au where all three spacecraft in this constellation are approximately located. }
    \label{fig:tdoa_schematic}
\end{figure}

At the top level, we take radio spectrogram data from three individual spacecraft (for the event studied in this work, Parker Solar Probe/FIELDS/RFS, STEREO-A/WAVES and Wind/WAVES, figure \ref{fig:lofar_psp_typeIII}a-c) which are mutually separated by large enough distances (light travel time longer than several minutes and therefore longer than each instrument's time resolution). For each spacecraft we measure the ``onset time'' of the burst in each frequency channel. We define this as the first time stamp at which the radio flux reaches half its maximum value in that channel (an example flux-time profile for similar frequencies in the three spacecraft receivers is shown in figure \ref{fig:time-profile}). By measuring the onset rather than, for example the time of peak flux, we minimize the effects of refraction and scattering effects by measuring the radio waves which have taken the shortest path length from source to receiver, although as seen in \ref{fig:time-profile} the burst rise time can be a similar timescale to the instrument resolution of Wind.

Extracting these features from the spectrograms produces a frequency ($f$) vs. time ($t$) profile for each spacecraft describing the ``time of arrival'' of the radio burst at each frequency. The frequency values of these profiles are set by the different channels measured by each instrument. To derive relative time delays between pairs of spacecraft we need to cast this data to a common grid of frequencies. To accomplish this, we fit a smooth curve to each frequency time profile. we fit a second order polynomial function to the set of values ($1/f$, $t$), where we use the reciprocal of the frequency (wavenumber) since this results in a curve well approximated by a quadratic. We then resample each ($f$, $t$) profile at a common set of frequencies to derive time of arrivals which can be directly differenced between spacecraft. This fitting process and output is shown in figure \ref{fig:feature-extraction} which shows that in addition to allowing frequency interpretation, (1) the fitted curve follows the onset data very closely (within the instrument time resolution in most cases), (2) allows rejection of outliers in the onset detection especially at higher frequencies where the background is noisier, and (3) smoothly interpolates the Wind/WAVES (and to a lesser extent S/WAVES) data which has an extremely coarse time resolution. 

Following the feature extraction step, for each pair of spacecraft, we find the \textit{difference} in time of arrival as a function of frequency. At a given frequency, we can combine this derived time delay with a baseline defined by the vector separation of the relevant spacecraft to produce a single 2D hyperbola in the ecliptic plane along which the radio emission at that wavelength originates subject to the following two assumptions:

\begin{enumerate} \item Free streaming radio emission from source to receiver. We note that by extracting the burst onset we have attempted to minimize this error source, but it otherwise largely remains unquantified.
\item Near-ecliptic propagation of the burst.  By ``near-ecliptic'', the precise approximation we are making is as follows: We are assuming that the source trajectory and the three spacecraft where the measurements are made are mutually coplanar. There will therefore be small but non-zero errors coming from the differing heliographic latitudes of STEREO A and PSP, and any angular displacement between the true burst propagation direction and the ecliptic plane. Although a full estimate is non-trivial, the magnitude of these errors can be estimated by finding the perpendicular light travel time between each location and the ecliptic plane. For the studied time interval, we find the following values for the constellation of receivers : PSP=5.42s, Wind/L1=0.01s, STEREO A=-1.00s. For the burst, we estimate the off-ecliptic latitude from figure \ref{fig:lofar3d} as $\lesssim 4^o$. At the maximum radius of around $50$R$_\odot$ we obtain a light travel time of 8s. In summation, we find the error is sub-dominant to the instrument resolution.
\end{enumerate}

Obtaining such a hyperbola for two pairs of spacecraft (three spacecraft total) is sufficient to constrain the source position to a single point. Algorithmically, it results in two points, one sunward of the baseline and one anti-sunward, but we can rule out the anti-sunward source due to (1) the monotonically decreasing solar wind electron plasma density which prevents radio emission travelling inwards in the heliosphere, (2) simply that there are no credible alternatives to the canonical picture that type III bursts originate at the Sun. Thus for each frequency we produce a 2D source position in the ecliptic plane.

Further, we can estimate a conservative error region by utilizing the instrument resolution for each spectrogram measurement. For PSP/RFS, S/WAVES and Wind/Waves these are err= 7s, 38.05s and 60s respectively. We therefore allow that each onset time is accurate to $\pm err/2$. These errors are then summed pairwise to give an error in relative time of arrival between each pair. As noted above our fitted quadratic curves to the burst profile track the measured burst onset to within these error bars for the majority of frequency channels. The propagation of this error to the source position effectively blurs the hyperbolae to give finite thickness. Intersecting these blurred profiles with three independent spacecraft measurements results in a hexagonal source region whose aspect ratio depends on relative spacecraft position (note that while the centroid is uniquely determined by two time delay measurements from three spacecraft, the error region is further constrained by including all permutations (three time delay measurements) from the three spacecraft. This hexagon can be found either by forming a minimum area convex hull from the analytic intersections of the edges of the blurred hyperbolae as illustrated in figure \ref{fig:tdoa_schematic}, or more simply as done in the main text in figure \ref{fig:psp_sta_wind} by randomly sampling time delay measurements within the stated error bounds and populating the source positions in the ecliptic plane.

This source analysis method and error region formation process, similar to that presented by \citet[][]{Alcock2018}, is summed up here in figure \ref{fig:tdoa_schematic}. We show both an example configuration with STEREO-A, STEREO-B and Wind in 2013 to clearly illustrate how the hyperbolae intersect to create the hexagonal error region as well as the configuration for the presently studied event where the hyperbola intersection is stretched and narrowed due to the orbital geometry.

\bibliography{sample63}{}
\bibliographystyle{aasjournal}

\end{document}